\documentclass{article}
\usepackage{spconf,amsmath,epsfig}
\usepackage{color}
\usepackage{xcolor}
\usepackage{graphicx}
\usepackage{amsthm,amssymb}
\usepackage{multirow}
\usepackage{subcaption}

\title{Early Flood Warning Using Satellite-Derived Convective System and Precipitation Data - A Retrospective Case Study of Central Vietnam}
\name{T.-V. La$^{*}$, T. H. Nguyen$^{*}$, P. Matgen, and M. Chini\thanks{$^{*}$These authors contributed equally.}}
\address{Luxembourg Institute of Science and Technology, 4362 Esch-sur-Alzette, Luxembourg\\
}

\begin{document}
%
\maketitle
\begin{abstract}

This paper addresses the challenges of an early flood warning caused by complex convective systems (CSs), by using Low-Earth Orbit and Geostationary satellite data. We focus on a sequence of extreme events that took place in central Vietnam during October 2020, with a specific emphasis on the events leading up to the floods, i.e., those occurring before October 10$^\mathrm{th}$, 2020. In this critical phase, several hydrometeorological indicators could be identified thanks to an increasingly advanced and dense observation network composed of Earth Observation satellites, in particular those enabling the characterization and monitoring of a CS, in terms of low-temperature clouds and heavy rainfall. Himawari-8 images, both individually and in time-series, allow identifying and tracking convective clouds. This is complemented by the observation of heavy/violent rainfall through GPM IMERG data, as well as the detection of strong winds using radiometers and scatterometers. Collectively, these datasets, along with the estimated intensity and duration of the event from each source, form a comprehensive dataset detailing the intricate behaviors of CSs. All of these factors are significant contributors to the magnitude of flooding and the short-term dynamics anticipated in the studied region.
\end{abstract}
\section{Introduction}
\label{sec:intro}

Severe convective systems (CSs) are regularly observed in sub-tropical and tropical regions, namely the Gulf of Guinea, the Gulf of Mexico, as well as the whole Southeast Asia, Australia, Indian subcontinent, etc. They are typically associated with extreme weather events such as intense thunderstorms, heavy rainfall, and strong surface wind gusts. The development of a CS often brings heavy precipitation over a small area in a short period, leading to flash floods. These events pose significant hazards due to their rapid onset, not allowing sufficient time for warnings and evacuations. However, accurately nowcasting and forecasting CSs remains a major challenge for meteorologists due to their sudden ignition and fast development. During the last four decades, considerable progress has been made in addressing this challenge, thanks to the arrival of many advanced GEOstationary (GEO) satellites, including Meteosat, GOES, Himawari, and Gaofen, covering Europe/Africa, Americas, and Asia-Pacific, respectively. Nevertheless, a more comprehensive understanding of the intricate relationships between deep convective clouds and extreme weather events is essential to enhance the prediction of natural disasters such as destructive floods caused by heavy rainfall. 

Mesoscale CS tracking and identification are mainly based on the analyses of brightness temperature and precipitation. \cite{la2021grl,la2021convective,la2022different} indicated the strong relationship between deep convective clouds and strong surface winds via intense downdrafts, based on the collocation of GEO (Meteosat) and Low-Earth Orbit (LEO) Sentinel-1 and Aeolus satellite imagery. Furthermore, \cite{la2021convective,la2022different} showed that the high-resolution (HR) surface wind patterns, derived from Sentinel-1 (S1) images, moved (horizontally) in the same direction as the deep convective clouds detected by Meteosat. In particular, \cite{la2022different} discussed the relationship between strong surface winds, heavy rainfall, and deep convection through the collocation of S1 SAR, WindSat scatterometer, and Meteosat data. They concluded that that intense surface winds were followed by heavy rainfall associated with deep convection approximately 30 minutes later. Regarding the relationship between deep convective clouds, heavy rainfall, and floods, \cite{del2020connecting} utilized radar data to characterize convective storms and connected them to the flood events that occurred on the northwestern Mediterranean coast. The results showed that the area was mostly affected by shallow but efficient convection, associated with high rainfall rates, which tend to exceed infiltration capacities and/or saturate catchments quickly and thus lead to fast run-off and flooding. \cite{atiah2023mesoscale} investigated the contribution of mesoscale CSs to the selected flood cases in southern West Africa based on the analysis of rainfall rates derived from brightness temperature obtained from the NOAA Climate Data Record of Gridded Satellite Data. It indicated that floods in this area had between 31\% and 60\% of contribution from mesoscale CSs. 

This paper tackles the challenges of issuing an early flood warning based on the analyses of multi-source LEO (SAR, scatterometers, and radiometers) and GEO satellite data. The different sensors provide complementary observations regarding hydrometeorological factors contributing to a flood event. This study explores the correlation between deep convective clouds identified by GEO satellites, patterns of strong surface winds, and heavy rainfall observed by LEO satellites, all of which are associated with severe flooding. The findings are validated through the mapping of flood extents using SAR technology. A specific case study of the flood events from Oct. $3^\mathrm{rd}$ to $10^\mathrm{th}$, 2020 that occurred in the central region of Vietnam is carried out in this paper. As reported by the Red Cross \cite{IFRC}, this area experienced prolonged, heavy rains that caused severe and widespread flooding and landslides in eight provinces between Oct. $6^\mathrm{th}$ and $14^\mathrm{th}$, 2020. This was due to the combination of numerous weather systems, including Inter Tropical Convergence Zones (ICZ) combining with cold air, tropical storms, and typhoons. 

\section{Methodology}
\label{sec:method}
\subsection{Data selection and preparation}

\begin{figure}[h]
\centering
\includegraphics[width=0.7\linewidth]{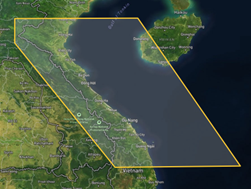}
\caption{Region of interest (ROI), including eight provinces and ocean areas of Vietnam.}
\label{fig:ROI}
\end{figure}

Fig. \ref{fig:ROI} illustrates the region of interest (ROI), i.e., $14^\circ-20^\circ$N, $103^\circ-110^\circ$E, including eight provinces of Vietnam, where flood events were observed starting from Oct. $6^\mathrm{th}$, 2020 \cite{VDDMA}. It also consists of the coastal areas from which CSs are formed and developed before arriving on land and producing strong surface winds and heavy rainfall. The ROI include the provinces Nghe An (NA), Ha Tinh (HT), Quang Binh (QB), Quang Tri (QT), Thua Thien Hue (TT), Quang Nam (QN1) and Quang Ngai (QN2), as well as the city of Da Nang (DN).

\begin{table}[h]
\caption{Satellite Images Used for Early Flood Warning - Case Study of the Central Region of Vietnam, Oct. 2020.}
\label{tab:data}
\centering
\scalebox{0.8}{
\begin{tabular}{cccc}
\hline
\textbf{Satellite} & \textbf{Type/Mode} & \textbf{Variable} & \textbf{Availability} \\\hline
\multirow{2}{*}{Sentinel-1} & LEO/ & Wind, & Oct. 03, 05, 06 \\
 &  C-SAR & Flood extent & \& 10 \\\hline
\multirow{2}{*}{WindSat} & LEO/ & Wind/  & Oct. 07 \& 08 \\
& Radiometer & Precipitation &  \\\hline
ASCAT- & LEO/ & \multirow{2}{*}{Wind} & Oct. 03-10\\
A/B/C & Scatterometer & & \\\hline
\multirow{2}{*}{SMAP} & LEO/ & \multirow{2}{*}{Wind} & Oct. 04, 06, 07  \\
& Radiometer & & \& 09 \\\hline
\multirow{2}{*}{Himawari-8} & \multirow{2}{*}{GEO} & Brightness  & Oct. 03-10, \\
 & & temperature & every 10 min \\\hline
\multirow{2}{*}{GPM IMERG} & \multirow{2}{*}{Multi-satellite} & \multirow{2}{*}{Precipitation} & Oct. 03-10, \\
 &  & & every 30 min \\
\hline
\end{tabular}}
\end{table}

\begin{figure}[h]
\centering
\includegraphics[width=0.7\linewidth]{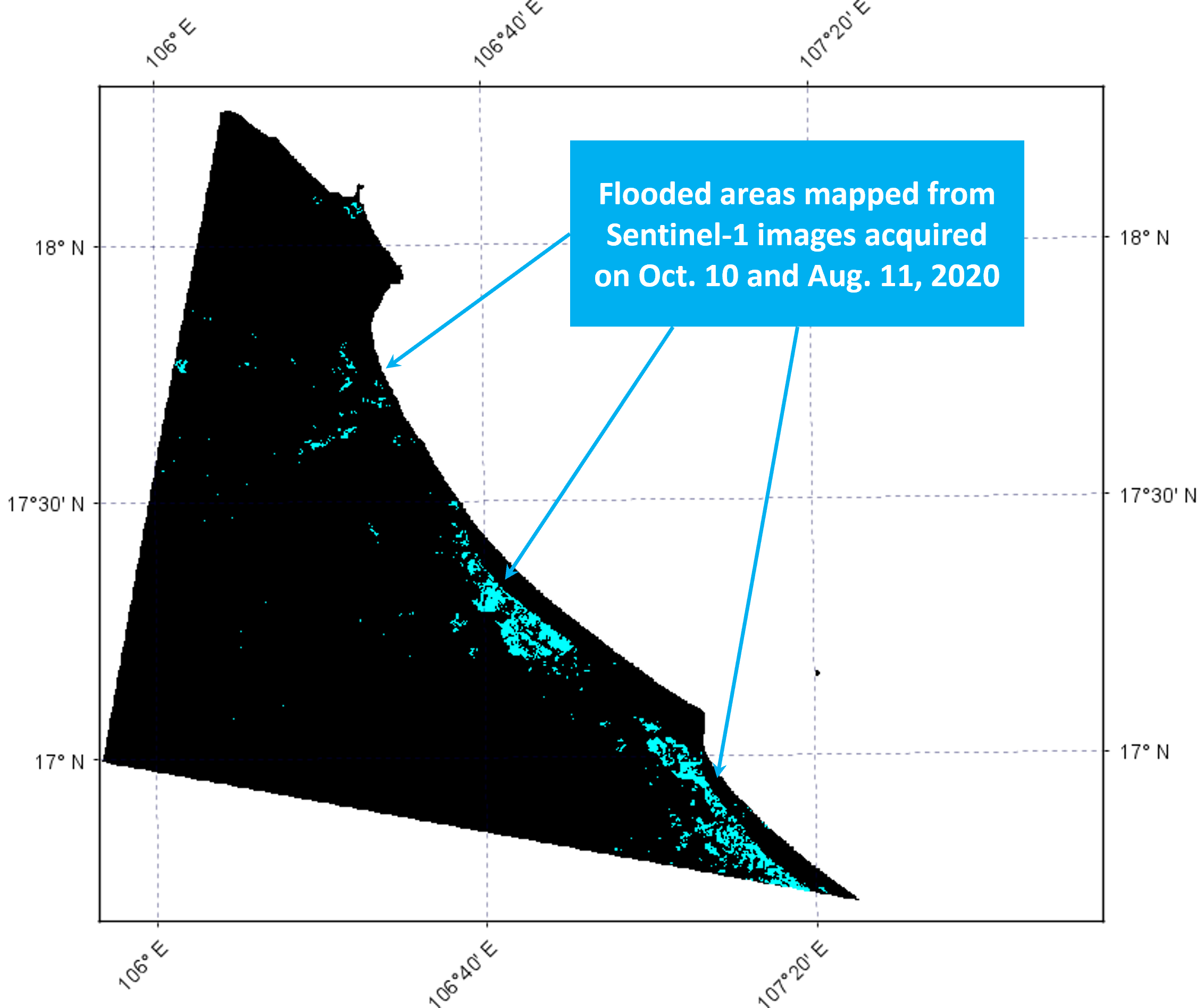}
\caption{S1-derived flooded areas observed along the coast of QB and QT provinces, Vietnam on Oct. 10$^\textnormal{th}$, 2020.}
\label{fig:floodmap}
\end{figure}

The satellite data, utilized to investigate the relationship between deep convective clouds, strong surface winds, heavy rainfall, and flood events for the period of Oct. 3$^\textnormal{rd}$ to 10$^\textnormal{th}$, 2020, are shown in Table~\ref{tab:data}. To detect deep convective clouds that can produce strong surface winds and heavy rainfall, we rely on the GEO Himawari-8 (H8) images provided every 10 minutes. We use S1 SAR images (available on four different days) to retrieve the HR surface wind speed and map flood extent. Fig.~\ref{fig:floodmap} presents the flooded areas observed along the coast of the QB and QT provinces, Vietnam, on Oct. 10$^\textnormal{th}$, 2020. They are mapped using the S1 image acquired on the Oct. 10$^\textnormal{th}$ with respect to a reference non-flood S1 image acquired on Aug. 11$^\textnormal{th}$, 2020. Recent research works \cite{nguyen2022,nguyen2023} on flood forecasting relied on S1-derived flood extent map as validation and assimilation data. The radiometers WindSat and SMAP also provide ocean surface wind speed and direction for 2 to 4 days, at a coarser resolution. Similarly, using scatterometer data (ASCAT-A/B/C), we can extract the wind speed over the ocean surfaces. WindSat data can also offer precipitation data in addition to half-hourly GPM IMERG precipitation accumulations \cite{meissner2017capability}.

\subsection{CS Identification from LEO and GEO Images}
\subsubsection{Deep convective clouds}
H8 images observe deep convective clouds that typically exhibit a low brightness temperature between 200 and 220 K. They correspond to convective wind patterns observed on LEO images, namely S1, WindSat, ASCAT-A/B/C, and SMAP.  The time-series of H8 images are also taken into account to detect cloud movements (i.e., general direction and velocity) during extreme weather, and track those with low temperature . This enables a broad assessment of the area affected by the investigated CS, particularly in the presence of concurrent heavy rainfall.

\subsubsection{Surface convective wind patterns}

Surface convective wind patterns can be derived from the images acquired by SARs, scatterometers, and radiometers. The wind speed observed over sea surfaces, potentially associated with the coldest convective clouds, can be estimated using S1 images. Such wind speed can be up to 25 m/s.  It has been shown that the intense downdrafts associated with deep convective clouds can induce surface wind gusts when they hit the surface of the sea \cite{la2021grl,la2021convective,la2022different}. In other words, vertical winds assert the active level of a CS, as well as being an element that induces heavier rainfall through downdrafts. The strong wind gusts significantly impact sea surface roughness, leading to observable high-intensity radar backscattering or areas with increased Normalized Radar Cross Section (NRCS) in the images captured by SARs and scatterometers. Using Geophysical Model Functions, e.g. CMOD for C-band, LMOD for L-band, and XMOD for X-band, we can retrieve surface wind speed from NRCS, as well as wind direction, and other radar parameters. 


Regarding the radiometer data, surface wind speed is retrieved from the sea surface brightness temperature measured by SMAP and WindSat. To this end, [8] took into account the differences between measured sea surface brightness temperatures and those of a flat ocean surface and matched these differences to the Radiative Transfer Model (RTM). Once the differences in brightness temperature and other parameters are determined, one can invert the RTM to estimate surface wind speed.

\subsubsection{Rainfall}

In addition to strong surface winds, deep convective clouds have the potential to generate substantial rainfall, a crucial factor contributing significantly to the initiation of flash flood events, as discussed in \cite{atiah2023mesoscale}. In this study, we utilized the precipitation accumulation data provided by the GPM IMERG [9] and by WindSat (over the ocean surface). As demonstrated in \cite{la2022different}, convective rainfall may occur subsequent to the onset of strong surface winds, aligning with the intensity of the deep convective clouds. 

\subsection{Data Fusion}
\begin{figure}[h]
\centering
\includegraphics[width=\linewidth]{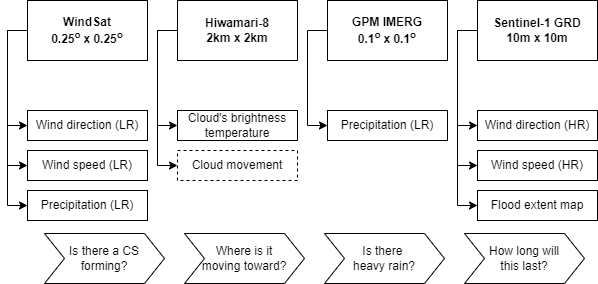}
\caption{Data and variables of interest.}
\label{fig:data}
\end{figure}


The combination of precipitation and brightness temperature has been leveraged to identify, locate, and track mesoscale CSs. Fig.~\ref{fig:data} summarizes this fusion framework between the input data. Using H8 and GPM IMERG, in addition to S1 HR and WindSat/ASCAT/SMAP low-resolution (LR) identification of surface wind direction and speed, a series of typical operational questions can be answered based on decision-level data fusion taking into account anecdotal evidence from each dataset. Except brightness temperature, every variable of interest can be estimated or measured from at least two satellites. This enables enhanced spatial coverage and temporal sampling, particularly for areas with restricted coverage or infrequent revisits. Surface wind speed estimates are provided in LR by WindSat/ASCAT/SMAP and in HR by S1 data. They are to be analyzed in conjunction with the location of low-temperature clouds and presence of intense precipitation, constituting deep convections. Precipitation measurements can also be either from GPM IMERG or from WindSat, offering complementary rainfall observations over ocean surface regions.

\section{Early Flood Warning Decision System}\label{sec:result}

This section unveils the outcomes of satellite data analysis and integration, focusing on the identification and surveillance of intense CSs and the corresponding extreme weather events responsible for the floods witnessed in the central region of Vietnam in October 2020. (Fig. \ref{fig:floodmap}). The identification of the relationship between deep convective clouds and extreme weather events is facilitated in recent years, thanks to GEO H8 data acquired every 10 minutes, and continuous observations every 30 minutes  from GPM IMERG Early Run. 

\begin{figure*}[t]
\centering
\begin{subfigure}[b]{0.24\linewidth}
 \includegraphics[width=5cm]{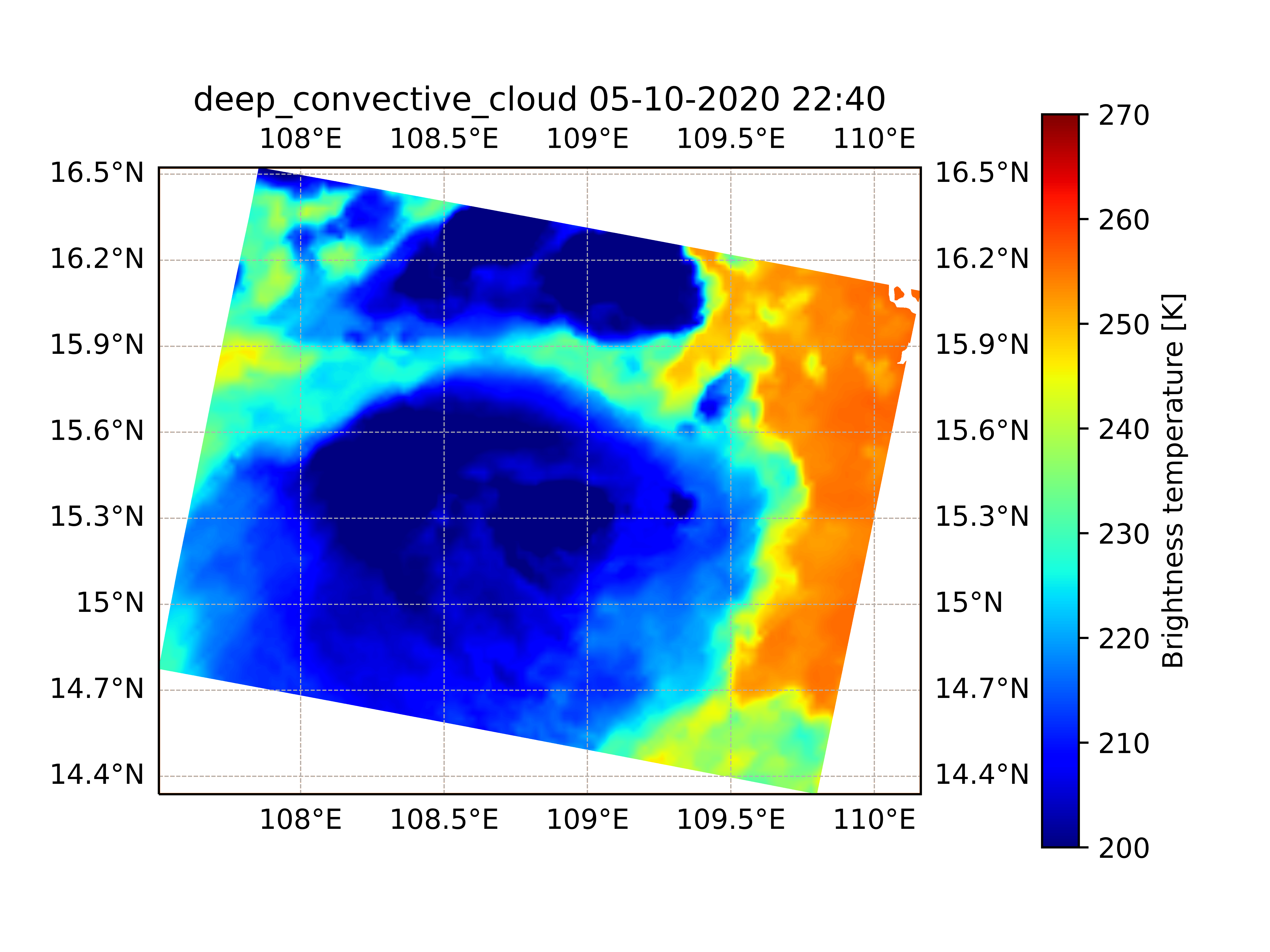}
  \caption{Himawari-8, 22:40 UTC}\label{sfig:4a}
\end{subfigure}
\hfill
\begin{subfigure}[b]{0.24\linewidth}
  \centering
  \includegraphics[width=5cm]{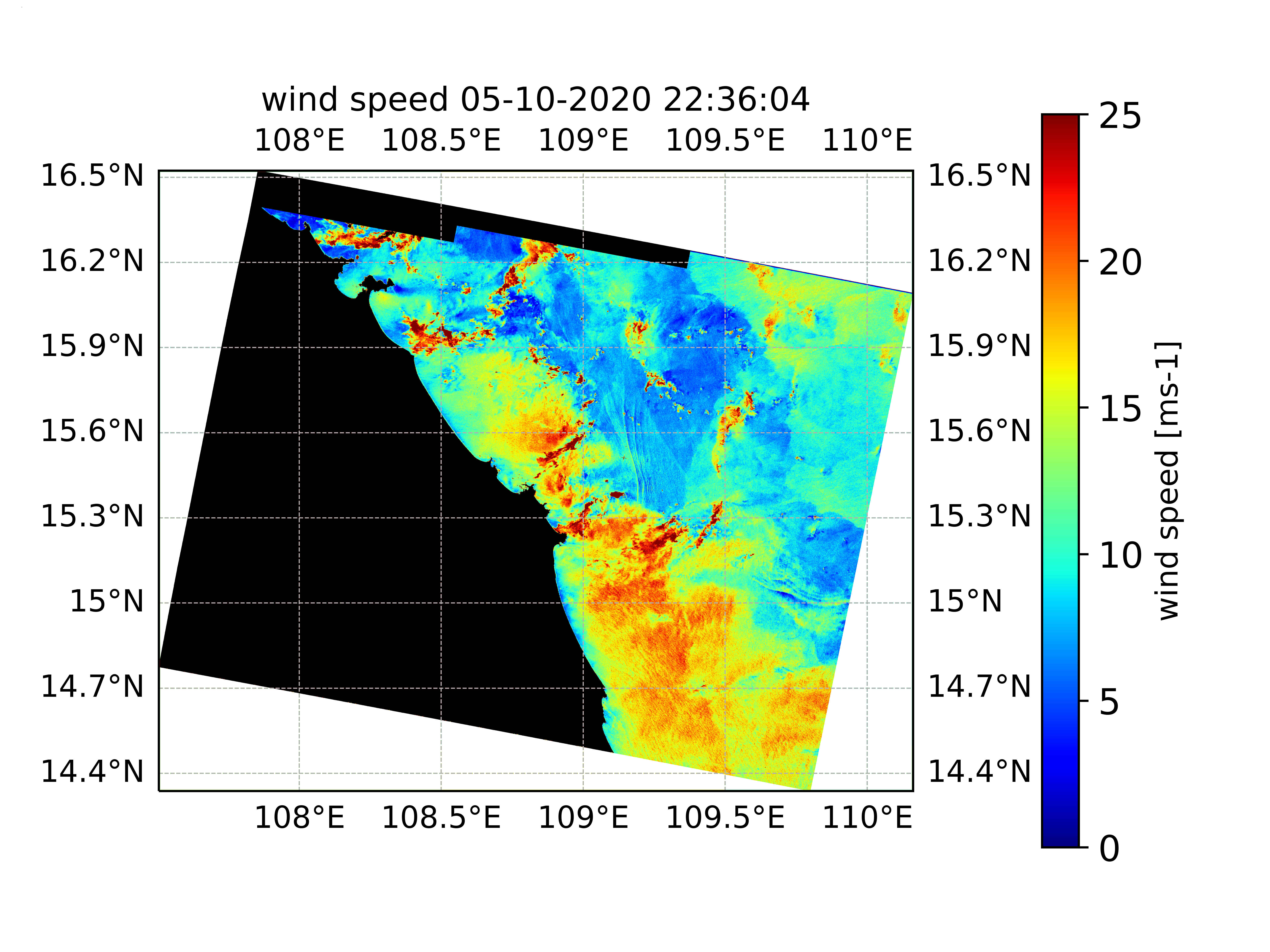}
  \caption{S1, 22:36:04 UTC}\label{sfig:4b}
\end{subfigure}
\hfill
\begin{subfigure}[b]{0.24\linewidth}
  \centering
  \includegraphics[width=5cm]{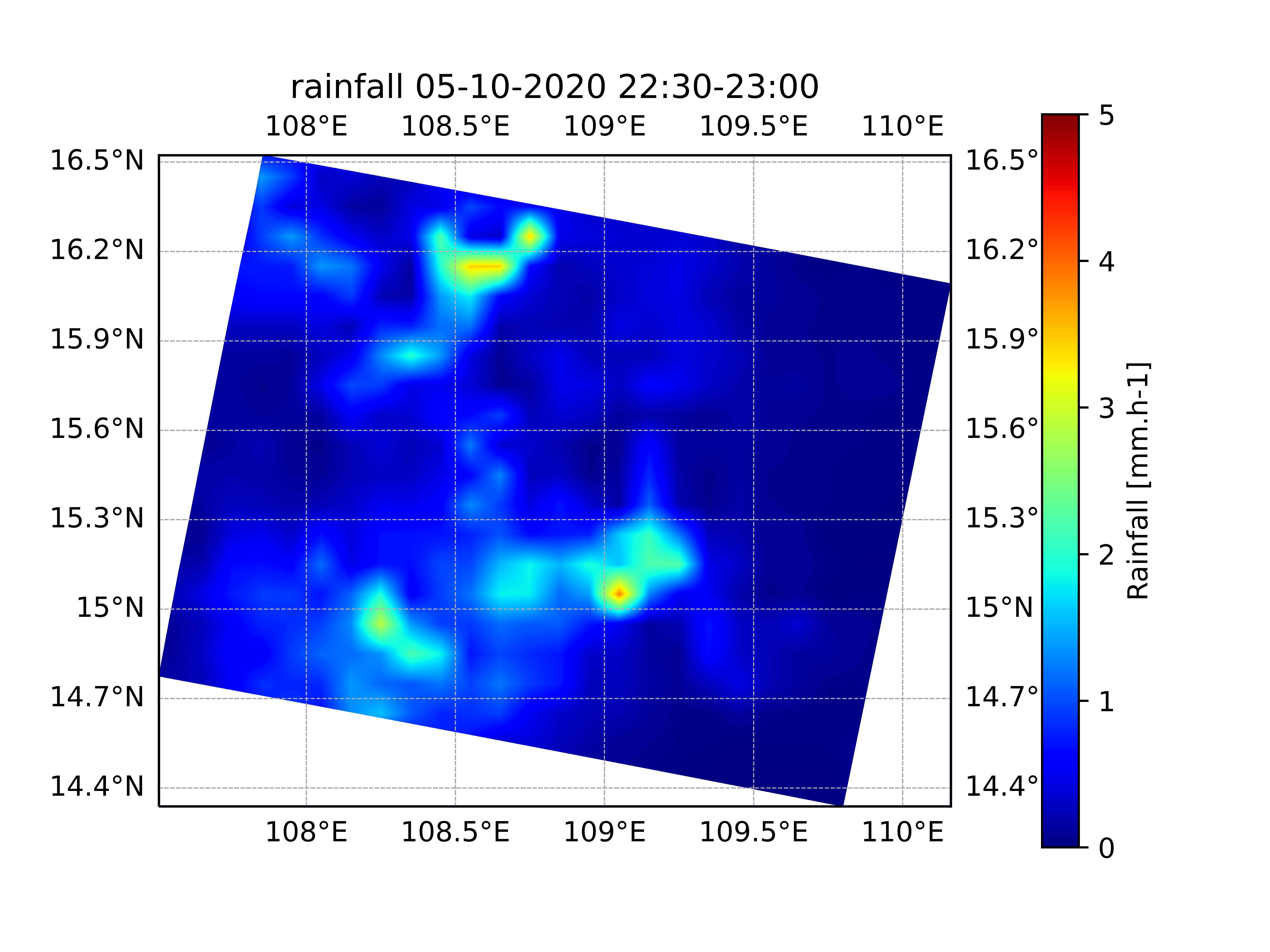}
  \caption{GPM, 22:30-23:00 UTC}\label{sfig:4c}
\end{subfigure}
\hfill
\begin{subfigure}[b]{0.24\linewidth}
  \centering
  \includegraphics[width=4cm]{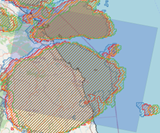}
  \caption{Extent of clouds} \label{sfig:4d}
\end{subfigure}
\caption{Combination of satellite data for observing a severe CS and its associated factors, on Oct. 5$^\textnormal{th}$, 2020, including (a) deep convective clouds detected by H8 band 10 (central wavelength: 7.35 $\mu$m), (b) strong surface wind pattern observed by S1 (c) heavy rainfall extracted from GPM IMERG, and (d) Extent of clouds under 220 K between 22:00-23:50 UTC showing a westward direction.}
\label{fig:Oct05}
\end{figure*}

Fig.~\ref{fig:Oct05} shows the collocation of H8, S1, and GPM IMERG images acquired on the Oct 5$^\mathrm{th}$, 2020, i.e., before the reported flood events \cite{VDDMA}. They show, respectively, (a) cloud brightness temperature, (b) strong surface wind patterns, and (c) heavy rainfall occurring in the region. The deepest (coldest) convective clouds depicted by a brightness temperature of around 200 K (Fig. \ref{sfig:4a}), align with strong surface wind zones ranging from 18 to 25 m/s along the coast (Fig. \ref{sfig:4b}), coupled with escalating rainfall accumulation (Fig. \ref{sfig:4c}). Such a combination can be found in convection lines, typically squalls \cite{la2021grl,la2021convective,la2022different} over the coastal areas. This substantiates the presence of a CS that developed along the coast (effectively an Intertropical Convergence Zone or ICZ), attributed to the S1-derived HR wind patterns and the frequent temporal sampling provided by H8 and GPM IMERG data. Next, considering the observed westward movement of these low-temperature clouds (Fig. \ref{sfig:4d}), jointly with the coarse surface wind direction provided by WindSat, it can be inferred that the area comprising four regions (namely TT, DN, QN1, and QN2) was poised to experience substantial rainfall, potentially leading to floods in the following days, with DN identified as the most vulnerable area. Using GPM IMERG data, the heavy precipitation persisted in these areas (partly due to the incoming tropical storm Linfa) until Oct. 9$^\textnormal{th}$, 2020, when it became less severe in DN, QN1 and QN2.

\begin{figure*}[!t]
\centering
\begin{subfigure}[b]{0.24\linewidth}
    \includegraphics[width=5cm]{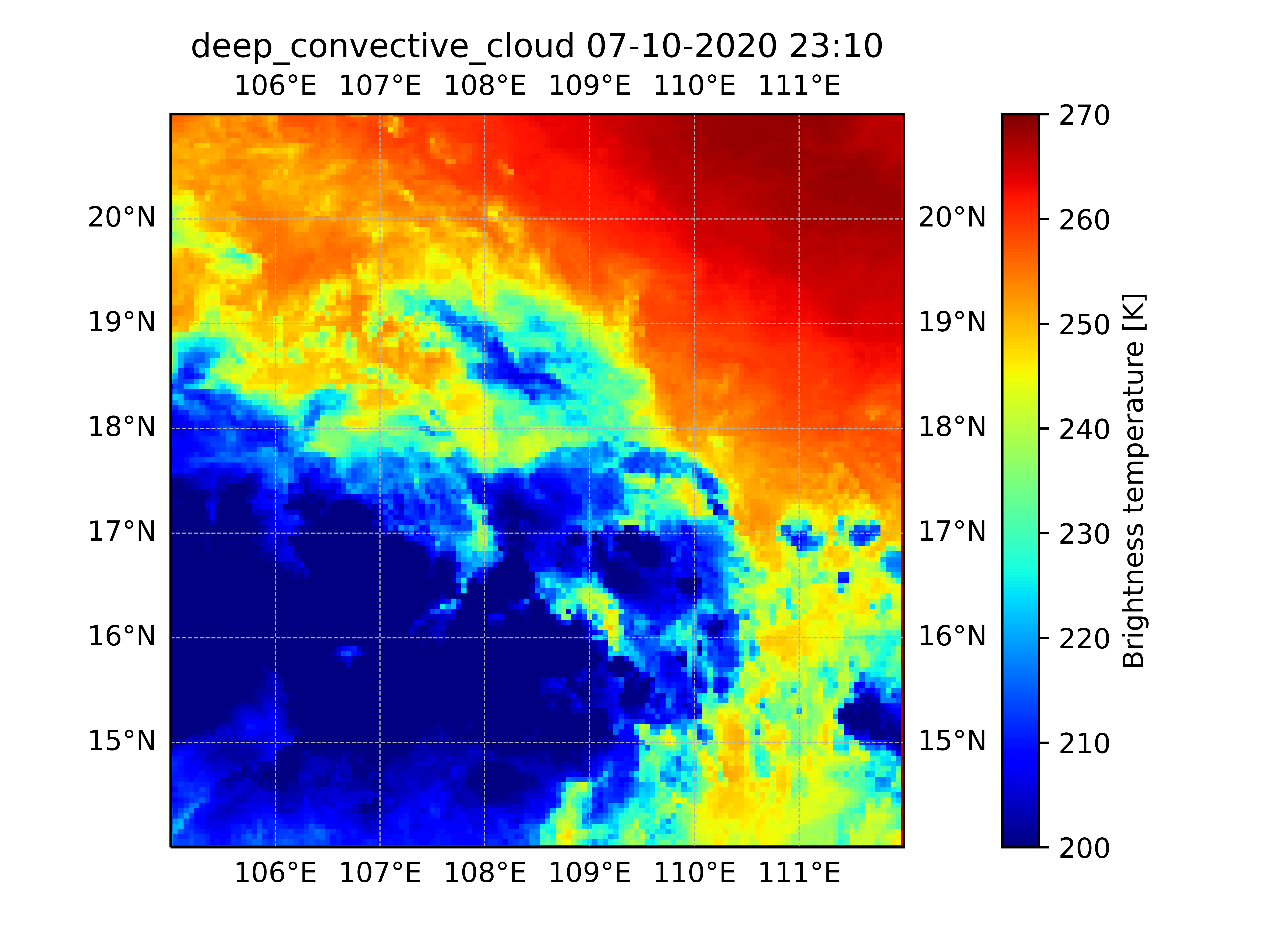}
    \caption{Himawari-8, 23:10 UTC}\label{sfig:5a}
\end{subfigure}
\hfill
\begin{subfigure}[b]{0.24\linewidth}
  \centering
  \includegraphics[width=5cm]{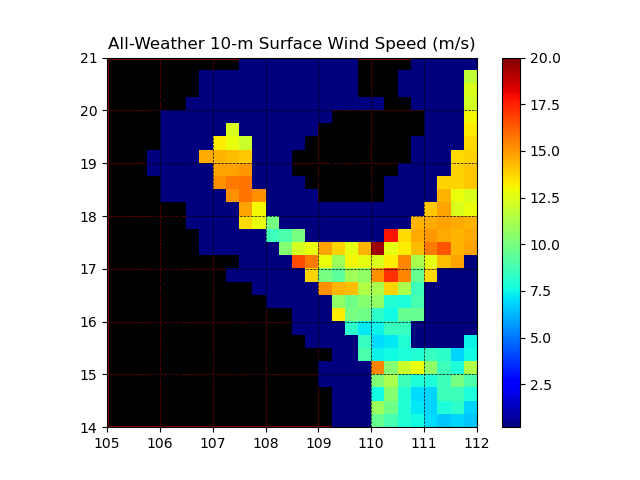}
  \caption{WindSat, 23:12 UTC}\label{sfig:5b}
\end{subfigure}
\hfill
\begin{subfigure}[b]{0.24\linewidth}
  \centering
  \includegraphics[width=5cm]{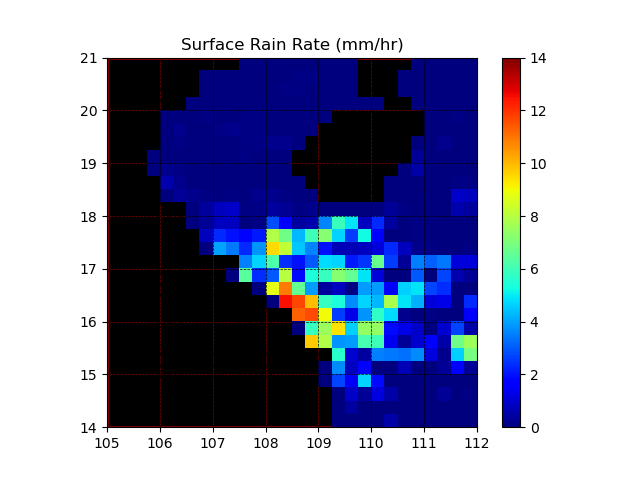}
  \caption{WindSat, 23:12 UTC}\label{sfig:5c}
\end{subfigure}
\hfill
\begin{subfigure}[b]{0.24\linewidth}
  \centering
  \includegraphics[width=5cm]{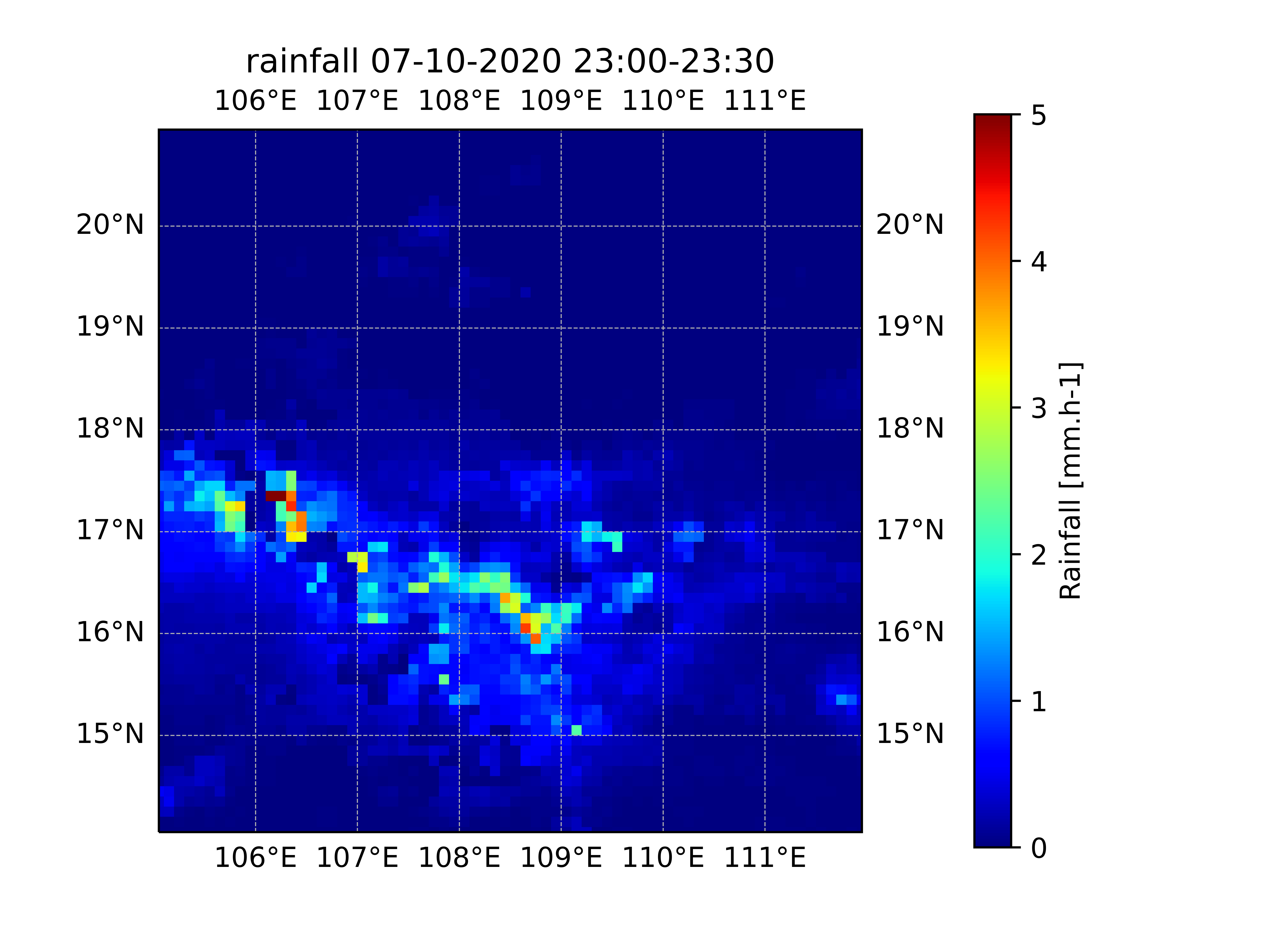}
  \caption{GPM, 23:00-23:30 UTC}\label{sfig:5d}
\end{subfigure}
\caption{Combination of various satellite data for observing a severe CS and its associated factors, on Oct. 7$^\textnormal{th}$, 2020: (a) deep convective clouds detected by H8 band 10 (central wavelength: 7.35 $\mu$m), (b) strong surface wind pattern and (c) heavy rainfall observed by WindSat, and (d) heavy rainfall extracted from GPM IMERG data.}
\label{fig:Oct07}
\end{figure*}

Fig. \ref{fig:Oct07} depicts the integration of H8, WindSat surface wind speed, and rainfall, along with GPM IMERG images obtained on October 7$^\textnormal{th}$, 2020. This timeframe falls after the occurrence of the initial ICZ and before the onset of the series of typhoons/tropical storms, which commenced on October 11$^\textnormal{th}$, 2020, as documented in \cite{IFRC,VDDMA}. They depict, respectively, observed deep convective clouds, strong surface winds, and intense precipitation. On the H8 image (Fig. \ref{sfig:5a}), as shown earlier in Fig. \ref{sfig:4a}, we can identify the deep convective clouds based on its low brightness temperature of approximately 200 K. It should be noted that like GPM IMERG data, WindSat can provide precipitation estimates over sea surfaces, but at a larger spatial resolution. Similarly, the WindSat-derived surface wind maps are coarser than S1 ones; however, their larger swath allow to observe a wider offshore wind area as well as identify future incoming events. 


At the same time and location, we observe strong surface wind patterns (12.5–20 m/s) and heavy rainfall rate (8–13 mm/h) on the WindSat images (Fig. \ref{sfig:5b}-\ref{sfig:5c}). GPM IMERG also confirms the intense precipitation present over the whole coastal region. Continuous observations in these areas, with H8, GPM and WindSat/ASCAT/SMAP between Oct. 7$^\textnormal{th}$ and Oct. 10$^\textnormal{th}$ suggests that these areas are likely to experience prolonged periods of intense winds and heavy rainfall. The matching of these observed variables demonstrates the strong association between deep convective clouds and extreme weather events. This further supports the widespread utilization of GEO images in combination with wind and rainfall data for the early detection of flood events in regions or seasons where severe CSs occur regularly. 

In addition, concerning wind speed data over the studied ROI, a summary of satellite observations leading up to  Oct 10$^\textnormal{th}$, 2020 combining S1, WindSat, ASCAT-A/B/C and SMAP data has been analyzed. In order to harmonize the wind speed estimated from SARs and scatterometers, they are categorized into “none”, “weak”, “moderate”, and “severe”, respectively corresponding to less than 5 m/s, 5-10 m/s, 10-15 m/s, and greater than 15 m/s. From this collection, it can be noted that on the Oct. 6$^\textnormal{th}$, 2020, there is an indication of wind getting stronger (starting from Oct. 5$^\textnormal{th}$ PM and subsequently reaching severe levels in the afternoon of October 6$^\textnormal{th}$. Over the next three days, severe winds continue to be prevalent across the entire ROI, In conjunction with the presence of observed low-temperature clouds and heavy rainfall resulting from the initial Intertropical Convergence Zone (ICZ) and tropical storm Linfa, it is evident that factors contributing to flooding have accumulated, even though official warnings had not been issued at that time. As a matter of fact, the areas impacted by heavy rainfall associated with many observed severe CSs correspond to the detected flooded areas (Fig. \ref{fig:floodmap}).

\section{Conclusions}

This paper presents a preliminary concept of an early flood warning system that is based on the use of multiple satellite data, namely Low-Earth Orbit (LEO) imagery data and Geostationary (GEO) imagery data. We examine a series of extreme events that transpired in central Vietnam in October 2020, with a specific focus on the period leading up to it (before Oct. $10^\mathrm{th}$, 2020). Within these events, various hydrometeorological indicators could be identified in advance, particularly those capable of tracking and monitoring extreme events, such as low-temperature clouds and prolonged heavy rainfall in specific areas. If these pertinent indicators had been provided in a timely manner, they could have been valuable for forecasting the geographical extent and severity of the ensuing extreme flood events Therefore, in this research work, we investigated the collocation of different LEO/GEO datasets in order to determine the causality between intense rainfall, low-temperature clouds, strong wind speed, and the potential for causing substantial flooding in the same location. This work demonstrates the feasibility of issuing a warning between several hours and one day in advance due to the presence of high rainfall rates and low-temperature clouds covering large coastal regions. In addition, utilizing H8 time-series of brightness temperature images, it becomes viable to forecast the trajectory and velocity of these convective low-temperature clouds moving from the sea towards the land. This allows us to identify the regions susceptible to significant rainfall. Lastly, using different low-resolution scatterometers in combination with the HR wind speed estimation from S1 images, local wind patterns and CS effects could be derived, showing the severity of wind in the studied areas.

\section{Acknowledgment}
This work is supported by the Luxembourg National Research Fund (FNR) in the framework of the CORE project C20/SR114703579.

\bibliographystyle{IEEEtran}
\bibliography{refs.bib} 

\end{document}